# ZKPs: Does This Make The Cut?

## Recent Advances and Success of Zero-Knowledge Security Protocols

Stavros Kassaras[1], Leandros Maglaras[1,2]

[1]Department of Computer Science and Computational Biomedicine, School of Science, University of Thessaly, Lamia, Greece, `kassaras.stavros@gmail.com`
[2]School of Computer Science and Informatics, De Montfort University, Leicester, UK

**Abstract.** How someone can get health insurance without sharing his health information? How you can get a loan without disclosing your credit score? There is a method to certify certain attributes of various data, either this is health metrics or finance information, without revealing the data itself or any other kind of personal data. This method is known as "zero-knowledge proofs".

Zero-Knowledge techniques are mathematical methods used to verify things without sharing or revealing underlying data. Zero-Knowledge protocols have vast applications from simple identity schemes and blockchains to defense research programs and nuclear arms control.

## 1 Introduction

Zero-Knowledge proofs (also commonly referred to as ZKPs) are used as security protocols through which a digital authentication process can be facilitated without the use of any passwords or other sensitive data. As a result of this, no information, either from the sender's or receiver's end can be compromised in any way. The idea underlying zero-knowledge proofs first came to the fore back in 1985, when developers Shafi Goldwasser, Charles Rackoff and Silvio Micali [1] presented to the world the notion of "knowledge complexity" — a concept that served as a precursor to ZKPs. Simply put, ZKP is a probabilistic based verification method, which means it provides "fact-like statements" and "statements about personal knowledge" that can accumulate to show that the validity of an assertion is overwhelmingly probable. In other words, they don't prove something that simply revealing it would be sufficient, but rather verify the knowledge of it. Moreover, the assertion cannot be verified by a third party.

This is especially handy in various situations, e.g. when we want to spend money without revealing how much we used or in which currency it was spent. Besides money, private information such as date of birth, bank statements, transaction histories, education credentials are vitally important. Companies like Facebook and Google leverage from this data by using it to optimize their services to better appeal to you, and to re-sell it to other companies. It is prevalent now than ever before the need to maintain privacy in a data-based world.

The paper is organized as follows. We provide an overview of the methodical approach used, important literature review and relevant references for this paper in Section 2. In Section 3 we will dive deeper to the nuts and bolts of ZKPs and we will mention important theorems that drove its evolution. In Section 4 we present notable and the most recent use cases of zero-knowledge protocols. In Section 5 and subsequently in Section 6 we are discussing known threats and solutions of ZKP technologies. Finally, we discuss the current state of Zero-knowledge protocols in Section 7 and conclude in Section 8.

## 2        Methodology

Firstly, we approached from a theoretical point of view the rudiments of the ZKPs field. This served not only as a concise presentation of the principles underlining ZKPs but also gave us a historical aspect on how these protocols evolved. In this first step we consulted Rubinstein-Salzedo's "Cryptography" as a primer. This book is less sophisticated from others but manages to cover substantial topics in cryptography in an informal view. Silverman's "An Introduction to Mathematical Cryptography" offered a mathematical approach for ZKPs and notions surrounding their functionality such as Complexity Theory and Random Oracle Models. A detailed and descriptive approach would require a strong background both in cryptology and in abstract algebra.

Secondly, we searched mostly online articles and blogs on the current trends and technologies on ZKPs protocols since internet is the first place that breakthrough methods make headlines. We made sure to accompany all the new implementations appearing in this paper with the relevant citations. A contemporary approach is not only mandated because the field of ZK applications is still in early stage but also it serves as a guide for the aspiring cryptographer or mathematician.

## 3        The Principles of Zero-Knowledge Proofs

We will start this Chapter by presenting two simple examples that demonstrate the concept of zero-knowledge proof.

### 3.1        Ali Babaa's Cave (Example #1).

The first and most important example is inspired by a paper titled "How to Explain Zero-Knowledge Protocols to Your Children" [2].

Consider, for the sake of example, a cave consisting of a circular tunnel. Diametrically opposite to the entrance of this cave, there is a door which can only be opened by a password. Although this situation is probably not a real life scenario, it is quite useful in the display of the basic properties of ZKP. Now Peggy (also known as the prover of the statement) knows the password to this door, and she wants to prove this to Victor (also known as the verifier of the statement) without actually disclosing it to him. They set off to complete the task as follows:

Peggy goes into a random branch of the cave (that is, left or right). She does this without Victor knowing which branch she chose. Standing at the entrance of the cave, Victor calls out a random branch (again, either left or right), where he wants Peggy to come out from. Providing she really does know the password, she can obey Victor every time, using the door if necessary.

However, if Peggy did not know the password, then she would only be able to return by the named path if Victor were to give the name of the same path by which she had entered. Since Victor would choose left or right at random, she would have a 50% chance fooling Victor. If both (prover and verifier) were to repeat the above process several times, say 20 times, Peggy's chance of successfully anticipating all of Victor's requests would become vanishingly small.

Thus, if Peggy repeatedly appears at the exit Victor names, he can conclude that it is extremely probable that Peggy does, in fact, know the secret word.

### 3.2    Two Balls and the Color-Blind Friend (Example #2).

Another classic example used to demonstrate ZKP is the following [3].

Imagine your friend is color-blind and you have two balls: one red and one green, but otherwise identical in their shape and size. To your friend they seem completely identical and he is skeptical that they are actually distinguishable. You want to prove to him they are in fact differently colored, but nothing else, thus you do not reveal which one is the red and which is the green.

You give the two balls to your friend and he puts them behind his back. Next, he takes one of the balls and brings it out from behind his back and displays it. This ball is then placed behind his back again and then he chooses to reveal just one of the two balls, switching to the other ball with probability 50%. He will ask you, "Did I switch the ball?" This whole procedure is then repeated as often as necessary. He knows if he switched the ball because he did it himself, and you know if he did (because you can see the color) without revealing to him the actual color of the ball.

The above examples demonstrate an important subtle feature, that of zero-knowledge. Victor cannot convince anybody else of Peggy's knowledge about the password. If Victor were to create a transcript (e.g. videotape the whole process) that could potentially be that of the communication between him and Peggy, it would be useless. This transcript would be indistinguishable from a transcript that is entirely fabricated by a cheating verifier. An outsider, watching the recording, could argue that Peggy and Victor agreed in advance about the sequence of chosen branches. Thus, such a recording will certainly never be convincing to anyone but the original participants.

### 3.3 Characteristics of ZKPs

**Interactive ZKPs**: The examples above are forms of interactive proofs since the prover, performed a series of actions to convince the verifier, of a certain fact. The problem with interactive proofs is their limited transferability: to prove an ability attribute, or possession of secret data to someone else, or to the verifier several times, the prover will have to repeat the entire process.

Interactive ZKPs have further properties, namely Completeness and Soundness.

*Completeness Property*: An interactive proof protocol is complete if, given an honest prover and an honest verifier, the protocol succeeds with overwhelming probability (i.e., the verifier accepts the prover's claim).

*Soundness Property[1]*: If the prover is lying, then he cannot convince the verifier that he is telling the truth, except with some very small probability.

*Zero Knowledge Property*: Let's consider the cave example once again and the videotape (transcript) Victor made. We mentioned that if the recording were to be seen by a third party, it would not convince this party for Peggy's knowledge. Thus, if Victor wants to convince a third party, he can ask Peggy to demonstrate the transcript once again, but this time it will be the third party who will pick his own sequence of challenges for Peggy (not Victor's challenges). If there is a way to forge a proof that is indistinguishable from a genuine one (as in the case of the videotape), we say that there is a simulator for the proof in question. A proof of knowledge has the zero knowledge property if there exists a simulator for the proof.

**Non-interactive[2] ZKPs (NIZK)**: In a non-interactive proof, the prover can deliver a proof that anyone can verify for themselves. This relies on the verifier picking a random challenge for the prover to solve.

Cryptographers Fiat and Shamir [4-5] found that an interactive protocol can be converted into a non-interactive[3] one using a hash function to pick the challenge (without any interaction with the verifier). Repeated interaction between the prover and verifier becomes unnecessary, since the proof exists in a single message sent from prover to verifier.

---

[1]   Soundness can be described mathematically as an expected polynomial time algorithm M with the following property: if a dishonest prover can with non-negligible probability execute the protocol with the verifier, then M can be used to extract from this prover the knowledge which with overwhelming probability allows subsequent protocol executions.

[2]   A Non-interactive, zero-knowledge proof example can be found here `https://blog.goodaudience.com/understanding-zero-knowledge-proofs-through-simple-examples-df673f796d99`

[3]   Only languages in BPP have NIZK proof systems, under suitable hardness assumptions, NIZKs exist for all languages in NP.

# 4    Applications

## 4.1    Blockchain Use Cases

So far ZKPs may seem like a conundrum and you might wonder if there are real world consequences. In this chapter we will present some important applications.

Probably, the most prominent use case of Zero-knowledge proofs is within the context of a blockchain ecosystem. It offers a lot of benefits in regard to validating cryptocurrency transactions without disclosing any data related to it - such as where the transactions originated from, where it went or how much money was transferred.

A real-world use case of this technology is Zcash, a crypto platform that employs a special iteration of zero-knowledge proofs - called zk-SNARKs - that allow native transactions to remain fully encrypted on the blockchain while still being verified under the network's consensus rules.

The possibilities of zk-SNARKS are impressive, you can verify the correctness of computations without having to execute them and you will not even learn what was executed - just that it was done correctly. zk-SNARK stands for Zero-Knowledge Succinct Non-interactive ARguments of Knowledge [6], "Zero-knowledge" because they don't reveal any knowledge to the verifier apart from ensuring that the transaction is valid, "Succinct" because the size of the proof is small enough to be verified in a few milliseconds, "Non-interactive" because the proof consists of a single message sent from prover to verifier and "Arguments" because the Soundness Property holds true.

The usual cryptocurrencies, like Bitcoin, validate their transactions by linking the sender and receiver addresses, and input and output values on the public blockchain. Instead of exposing the above components, Zcash make use of zk-SNARKs to "obfuscate" them - it diminishes any meaningful connection between sender, receiver and amount.

In a nutshell, if a sender wants to create a shielded transaction, he constructs a proof to show that with high probability, the input values sum to the output values for each shielded transfer, the sender proves that they have the private spending keys of the input notes, giving them the authority to spend and the private spending keys of the input notes are cryptographically linked to a signature over the whole transaction, in such a way that the transaction cannot be modified by a party who did not know these private keys.

Users of cryptocurrencies often couple them with network-layer privacy enhancements like Tor to level up their anonymity (we should better say pseudonymity in this case) with unpleasant results despite their efforts [7], unlikely Zcash does not suffer from the same threats.

## 4.2    From zk-SNARKs to zk-STARKs

As this wasn't enough there is a more developed version of zk-SNARKs, it is called zk-STARKs - Zero-Knowledge Scalable Transparent Argument of Knowledge and it was introduced in 2018 (very recently) by Eli Ben [8].

Prior to the creation of zk-STARKs, zk-SNARKs required a trusted party to initially setup the ZK proof system which introduced the vulnerability of those trusted parties compromising the privacy of the entire system (read more about vulnerabilities in Section 5). zk-STARKs improve upon this technology by removing the need for a trusted setup. In other terms, zk-STARK proofs present a simpler structure in terms of cryptographic assumptions.

The great advantage of zk-STARK is its scalability, meaning it can move computations and storage off-chain. Off-chain services will be able to generate STARK proofs that attest the integrity of off-chain computations and then can be integrated back on chain for any interested party to validate the computation. Also, while zk-SNARK communication complexity - that is the amount of communication needed to solve a problem distributed among two or more parties - increases in a linear fashion, on the other side zk-STARK develops in the opposite way, and increase only slightly as the computation size grows. The same applies to the verifier complexity. zk-STARKs are about 10 times faster than zk-SNARKs as computation size increases.

### 4.3 Quantum Resistant

Quantum computing has become a topic of interest and despite the fact they are characterized with many novel attributes, unfortunately the truth is far from reality. Quantum computers can achieve only special kinds of calculations and some of them could exploit today's cryptographic techniques. Encryption schemes based on RSA and Elliptic Curve Cryptography could prove obsolete in the near future. Notice that these algorithms rely on private and public key pairings, something that doesn't apply in the case not only of zk-STARKS but also other ZKP methods in general.

### 4.4 Zero-knowledge proofs in Banking

In October 15 in 2018, ING published a report and subsequently an article [9], announcing its own addition of ZKP to the blockchain technology. ING Bank is continuing further down the path of advanced blockchain privacy with the release of its Zero-Knowledge Range Proofs (ZKRP) solution.

The ZKRP scheme proposed can be used to prove a number is within a specific range. For example, a mortgage or loan applicant could prove that their salary or credit score sits within a certain range without revealing the exact figure. As such range proofs are computationally lighter than regular zero-knowledge proofs and run faster on a blockchain.

Not long after this, ING took the solution a step further and introduced Zero-Knowledge Set Membership (ZKSM), described in [10], going beyond numerical data to include other types of information, like locations and names. For instance, banks could validate that a new client lives in a country that belongs to the European Union, without revealing the country. Simply put, this technology allows information to be shared without revealing contextual details.

### 4.5    Nuclear Disarmament Applcations

Zero-knowledge methods have been devised originally for computational tasks but recently translated into use for a physical system. At the Department of Energy's (DOE) Princeton Plasma Physics Laboratory (PPPL) researchers have developed an experimental verification protocol for weapon dismantlement agreements [11]. Their method includes a system that can compare physical objects while potentially protecting sensitive information about the objects themselves.

The process to prove two objects are identical - potentially nuclear warheads – is as follows: the prover provides two radiographic films already exposed with the inverse image of one test object and place them in two individual sealed boxes. The objects are placed in front of the boxes and getting radiated. This operation is essentially equivalent to adding a positive image on top of a negative image. The verifiers accepts the proof if both images after radiated are flat gray – meaning there is a 50 percent probability that the objects are indeed identical - otherwise he/she  rejects it. If multiple tests are run simultaneously and the inverse images are randomized to the transmission pattern of the test object, the probability that they are not identical falls even more. The proof is zero-knowledge because the verifier does not learn anything beyond the result of the proof.

### 4.6    Other Use Cases of ZKPs

- Ethereum: Ethereum is also working with zk-SNARK proofs since its Byzantium update in 2017 [12].
- PIVX: PIVX is a Proof of Stake (PoS) blockchain-based cryptocurrency created in 2016. At its core, it relies on fungibility, transaction privacy, and community governance. PIVX utilize zero knowledge proofs via ZeroCoin protocols [13].

### 4.7    Possible Use Cases of ZKPs

There are many areas that can be enhanced by using Zero-Knowledge protocols where trust is required and there are large incentives to cheat, such as:

**Chain Voting Models**: Voting can refer either to political elections or corporate voting, where shareholder participation is a longstanding economical pillar. In any voting procedure, security, anonymity and trust are of paramount importance since these parts are most likely to fail and participation might be lower in the possibility of censorship.

These issues can be resolved with a zero-knowledge method. The whole procedure can move on a public blockchain. Every eligible voter (or shareholder) can cast their ballot without revealing their identity and even more they can ask for verification of their vote to ensure their ballot was counted.

**Running a computation and verifying its results**: In the last years there is a trend for research centers and enterprises to migrate their data to outside providers. This practice raise concerns about the integrity and confidentiality of computations conducted on this data. We can imagine for example, a research medical center who wishes to have access in a private data center that contains genetic information of individuals, in order to apply a new formula.

This begs the following question, how we can verify the computations of the formula and at the same time achieve it without disclosing patients' identity? A Zero-knowledge method could answer both of these contradictory objects. The data center can apply on behalf of the researchers the formula and prove them in zero-knowledge the result without compromising the individuals' confidentiality [14].

**Data Auditing**: More and more users and enterprises resort in data centers due to storage limitations or for specialized services. Data integrity and availability is of major concern for cloud storage services while users uploading their personal data together with authentication information. This means users have no longer possession of their data that may face risks like loss, corruption or the purchase from a third service or company. The same concerns apply also to distributed ledgers.

Auditing from a third party is critical to prove data centers, financial institutions and exchanges are complying with regulations like GDPR. Profiting from zero-knowledge methods, we can construct protocol schemes to prevent the leakage of verified and private data, a problem that existing auditing protocols suffer from and can have devastating effects. The reader can find recent research in [15-16].

## 5 Threats and Vulnerabilities

Zero-knowledge proofs are by definition methods which satisfy the appropriate security features that interest us. Attacks and vulnerabilities could be found in the designing of a ZKP protocol or in the system resources that support the realization of such features. The former one is most likely to happen since such designing requires a higher level of technical and theoretical sophistication that in the process a mathematical mistake might occur. Such mistakes could go completely unobservable like the "Infinite Counterfeiting Vulnerability[4]", a mathematical false in a research zk-SNARKs paper that could irreparably damage the market since other cryptocurrencies employ the same algorithms.

---

[4] The story behind it is very interesting and can be found here `https://dci.mit.edu/research/2019/2/6/dci-director-was-interviewed-for-fortunes-latest-article-zcash-discloses-vulnerability-that-could-have-allowed-infinite-counterfeit-cryptocurrency`

### 5.1 Parameter Problem

zk-SNARKs have undergone significant tinkering and exploration into their real-world application and efficiency improvements, yet is not a perfect cryptographic tool.

For the zk-SNARKs to work, an initial setup phase is required where the so called "system parameters" are generated who act as a common reference string shared between prover and verifier and need to be built-in in advanced in every zk-SNARK implementation. This process is known as a "trusted setup" which is a highly polarizing event. If the parameters get compromised, a malicious user could use it to generate fake proofs and theoretically create infinite amounts of counterfeit coins of the native token (not only for Zcash but for any other crypto currency that adopts the zk-SNARK technology) without anyone knowing.

We can imagine a hypothetical scenario where elections take place on a blockchain using zk-SNARKs for proving votes. A powerful entity – such as a politician in the case of a Presidential vote – would have strong incentives to involve the parties with the setup of the zk-SNARK system to share the setup parameters. The entity could tilt the election in their intended direction by creating false proofs for votes taking place.

The biggest problem with the zk-SNARK approach is that users need to implicitly trust in the setup phase and the parties involved to setup the system honestly. Users of the system will never actually know if the setup phase was compromised at the point of setup, or at some point in the future. In other words we could say that the system is as much secure as the incentives to circumvent the system are low. So, if this is the case, the door remains open for a system where users do not implicitly need to trust the parties involved in the system's setup to be honest.

### 5.2 Possible Attacks

While cryptocurrencies increase their use as an actual currency and payment method so the interest for Simple Payment Verification (SPV) increases to support users who cannot hold the full blockchain ledger in their mobile devices. As cryptocurrencies gear towards portability, software designers have not given much attention to system integrity issues, thus system vulnerabilities are often unobservable. One of these dangers are fault attacks[5], which potentially could extract data from the CPU or memory of a device at the time a ZKP protocol is taking place.

In [17] the authors demonstrate the first (at the time) fault attack which is initiated purely form software – it removes entirely the physical access. A malicious user could unleash various vector attacks or a combination of them to induce faults in a CPU core while a ZK protocol is under verification. An attacker could fault the steps - and the respective data - which a verifier and a prover follow in a ZK protocol, and intercept (modify or destroy) the messages form one another.

---

[5] A fault attack is an attack on a physical electronic device which consists in stressing the device by an external mean in order to generates errors in such a way that these errors leads to a security failure of the system.

# 6 Solutions

One way the Zcash team got around the parameter problem was to create a multi-party computation ceremony [18-19]. During the parameter creation a set of random numbers are produced, a process similar to the setup phase of a public-key cryptosystem. Zcash team refers to these random numbers as "toxic waste" to emphasize the need to get disposed with extreme caution. Multiple independent parties involved collaboratively in the construction of the parameters. It is apparent that it takes all the participants of the "ceremony" to be compromised or be dishonest in order to give away the parameters.

Another way to remedy the parameter problem is the choice users have to send tokens privately or publicly. Zcash concerning, the privacy features are not obligated, but are rather customizable.

More powerful constructions of zk-SNARKS (and generaly non-interactive ZKPs) can rise to the challenge of the parameter setup. One of these, already mention above, is zk-STARKS. Another one is zk-ConSNARK which is developed by Suterusu[6]. Suterusu integrates a state of the art zero-knowledge schemes which are scalable and free of complex multi-party setup protocols. ConSNARKs can produce a sound blockchain ecosystem, absence of manipulation for the users and massively improved efficiency.

# 7 Discussion

We start our discussion by noticing first some technical details. The maximum rate, at which a blockchain protocol/technology is processed, is determined by the size of it, the size of the transaction and the intricacy of the underlying computations, consequently this is determined by the ZKP protocol/technology the blockchain adopts.

For example the complexity in terms of size of zk-STARKs rises much slower that the zk-SNARKs for one-time setup, after this phase SNARKs have much less size needed for computation in verifying the proof in faster times. On the other hand zk-ConSNARKs allows for very small and constant size proof computations, which leads to much faster generation and verification times. We could argue that as zero-knowledge technologies evolve, they push for smaller and smaller and/or constant size of proofs that can succeed better verification times [20].

**Table 1.** Comparison of three major ZKPs. Source: Adapted from [21].

|  | Proof Size | Prover Time | Verification Time |
|---|---|---|---|
| SNARKs | 288 bytes | 2.3s | 10ms |
| STARKs | 45KB-200KB | 1.6s | 16ms |
| Bulletproofs | ~1.3KB | 30s | 1100ms |

The above table presents the comparison of three major ZKPs technologies. We observe the addition of Bulletproofs [22], this protocol falls under the "range proof" style (as ING's ZKRP). STARKs are faster than SNARKs which are faster than Bulletproofs, but as we go in the opposite direction proofs shrink in size. Bulletproofs are much shorter in size than STARKs, a property of much importance, even if we go about small differences in the order of hundreds of bytes (from SNARKs) to the order of hundreds of kilobytes proof (that of STARKs), it might be a "killer" factor. It seems like there is an inherent trade-off between size and speed. This is not the only trade off. SNARK is the only protocol that requires a trusted setup, but verifying it is less time consuming than Bullerproofs. The same goes for STARKs which do not need a trusted setup. We could say that we trade confidetiality for efficient validation.

In respect to privacy and confidentiality, most zero-knowledge technologies adopt zk-SNARKS implementations but all require a trusted setup event, meaning it is a one-time event, if a vulnerability or mathematical mistake is to be found, a whole new multi-party ceremony needs to be deployed which is an extremely complex procedure. There are few zero-knowledge breakthroughs that have got ridden the obligation of an initial setup.

We can argue, in the context of a zero-knowledge protocol embedded in a system, that one rule is applied; the system is only as good as the secret it is trying to conceal.

## 8    Does It Make The Cut? (Conclusions)

So much of our world is dominated by services that gobble up our personal information, they abuse these data, they sell it to the highest bidder, they do not protect in on their own servers and they essentially leaving it out for ransom. It is vital now more than ever before the need for privacy. Zero-Knowledge innovations, like zk-SNARKs, are up to the task of preserving the confidentiality and the security of users' data. They have the potential to enable trust levels that have never been achieved before. On the other side, since ZKPs have been theorized in 1985, it is the last six or so years we begun to use it in practice. We are still experimenting and try to understand how most effectively to apply it. It might take years until we manage to harness it true potential.

We still have to overcome many challenges and to observe a broader range of applications, but there is no doubt we hold in our hands a novel class of technology, one that sparks further development and innovation.